\documentclass[showpacs,pra,superscriptaddress,twocolumn]{revtex4-1}
\usepackage{amsfonts}
\usepackage{amsmath}
\usepackage{graphicx}
\usepackage{amssymb}
\usepackage[breaklinks=true,dvips]{hyperref}
\usepackage{breakurl}
\usepackage[utf8x]{inputenc}
\usepackage{mathrsfs}
\usepackage{color}

\hypersetup{
  colorlinks=true,linkcolor=blue,citecolor=blue,
  filecolor=blue,urlcolor=blue,breaklinks=true
}
\begin{document}

\title{Goldbach conjecture sequences in quantum mechanics}
\author{Thiago Prud\^encio}
\email{thprudencio@ufma.br}
\affiliation{Coordination in Science and Technology - CCCT, Federal University of Maranh\~ao - UFMA,
Campus Bacanga, 65080-805, S\~ao Lu\'is-MA, Brazil.}
\affiliation{
   Department of Physics, Federal University of Maranh\~ao - UFMA,
Campus Bacanga, 65085-580, São Lu\'is-MA, Brazil.
}

\author{Edilberto O. Silva}
\email{edilbertoo@gmail.com}
\affiliation{
   Department of Physics, Federal University of Maranh\~ao - UFMA,
Campus Bacanga, 65085-580, São Lu\'is-MA, Brazil.
}
%\date{\today }

\begin{abstract}
We show that there is a correspondence between Goldbach conjecture sequences (GCS) and 
expectation values of the number operator in Fock states. We demonstrate that depending on the normalization 
or not of Fock state superpositions, we have sequences that are equivalent and sequences that are not equivalent 
to GCS. We propose an algorithm where sequences equivalent to GCS can be derived in terms of expectation values 
with normalized states. Defining states whose projections generate GCS, we relate this problem to eigenstates of 
quantum harmonic oscillator and discuss Fock states directly 
associated to GCS, taking into account the hamiltonian spectrum and quantum vacuum fluctuations. Finally, we address the problems of 
degeneracy, maps associating GCS and Goldbach partitions. 
\end{abstract}

\pacs{03.65.Ta, 03.65.Fd, 42.50.Gy, 02.10.De}
\maketitle

\section{Introduction}

One of the simplest assertions in mathematics, proposed by Goldbach 
to Euler in 1742, the Goldbach's
conjecture (GC), also called strong GC, establishes in its simplest form that even
numbers greater than $2$ are always written as a sum of two prime numbers. If we go from the first term, $4=2+2$, to a given 
intenger $k$, that corresponds to a term of type 
$2(k+2)$, we have finite GC sequences (GCS). It is an open problem to prove that such GC is valid for an arbitrary sequence, i.e., 
for any integer $k\geq 0$, although there was not found any exception \cite{yuan,dox,languasco}. The main point in this problem 
is dealing with an infinity of even numbers and a corresponding infinity of prime numbers, i.e., to prove that there is no exception for 
any even number. From the fundamental theorem of arithmetic, any natural number $n$ can be written as a product of primes $p_{1}...p_{k}$, for a given $k$,
 $n=p_{1}...p_{k}$ and, in particular, it is also 
proved that there is an infinity of primes, since $n+1$ cannot be completelly divided by $p_{1}...p_{k}$, i.e, there is a 
remainder $1$. Since Cantor \cite{cantor}, we know that there are many types of infinity and that the infinity associated the 
sets of even numbers, natural numbers and prime numbers have the same cardinality $\aleph_{0}$. As in the example 
of other mathematical problems \cite{sierra,shor,sierra2,aq,lidar}, we know that there is advantage in writing the mathematical problem in a way 
to make use of the advances in quantum algorithms, what can be useful for a quantum computer processing. As such, we can reexpress the
GC problem quantum mechanically, looking for quantum algorithms to write such sequences in order to
achieve an arbitrarily large GCS. From the fundamental point of view the states associated to GCS can also 
be connected in important question of physical interest \cite{lozano}. 

In this letter, we consider the problem associated to the generation of GCS in quantum mechanics. 
We consider sequences involving Fock states and consider their equivalency to the GCS.
We show that there is a correspondence between the GCS and expectation values of the number operator in 
Fock states. From the experimental point of view, Fock states can be created in 
many different quantum systems \cite{sayrin,guerlin,lupascu,chun,pons,ss,hofheinz,wang,brune1,brune2,raimond,ralph,lu} 
and are, consequently, 
adequate for the problem in view. We demonstrate that depending on the normalization 
or not of the superpositions of Fock states, we can generate sequences that are equivalent and sequences that are 
not equivalent to the GCS. In order to circumvent the problem related to the normalization, we propose 
an algorithm to generate a sequence involving expectation values with normalized states that is equivalent to GCS. We also 
propose states associated to GCS that generate all GC terms by means of projection relations and consider their association to the 
Hamiltonian of a quantum harmonic oscillator. We show how these states are related to eigenstates of the harmonic oscillator and how 
the Hamiltonian acts on them. Considering Fock states directly associated to the GCS, we disscuss its role in the 
spectrum of the quantum harmonic oscillator and association to vacuum fluctuations. Finally, we address the degeneracy problems related to states 
associated to GCS.

\section{GCS in terms of expectation values}

We first restrict ourselves to Fock states $|k\rangle $, with integers $k\geq 0$, that fulfill an eigenvalue equation for the number
operator $\hat{N}=\hat{a}^{\dagger }\hat{a}$, given by $\hat{N}|k\rangle =k|k\rangle$. It is well known that these states 
can be generated from the vacuum $|0\rangle $ by means of repetitive actions of the 
creation operator $\hat{a}^{\dagger }$, 
\begin{eqnarray}
|k\rangle =(\sqrt{k!})^{-1}\left( \hat{a}^{\dagger }\right)^{k}|0\rangle,
\end{eqnarray}
and can be annihilated by successive actions of the annihilation operator, $\hat{a}$. The expectation 
value of the number operator in the pure the Fock state $|k\rangle$ is 
then just $\langle k|\hat{N}|k\rangle =k$. We are also interested in superpositions of Fock states, that 
can also be generated from the vacuum by means of two different 
actions of the creation operator in a superposition of two orthonormal states
\begin{eqnarray}
\frac{|k\rangle \pm |k^{\prime }\rangle }{\sqrt{2}}=
\frac{1}{\sqrt{2}}\left((\sqrt{k!})^{-1}\left( \hat{a}^{\dagger }\right)^{k} 
\pm (\sqrt{k'!})^{-1}\left( \hat{a}^{\dagger }\right) ^{k^{\prime }}\right)|0\rangle \nonumber \\
\end{eqnarray}
where $1/\sqrt{2}$ is due to the normalization factor.

A special type of superposition is $|k\rangle +
|k\rangle = 2|k\rangle$, involving the same Fock state. Taking into account the normalization, this state is equivalet to $|k\rangle$,
with the denormalization factor $2$. Usually, such superpositions are not a new state, in particular $|k\rangle$ 
and $2|k\rangle$ correspond to 
the same physical state.

In the case of the superpostion state $|k\rangle
+|k^{\prime }\rangle$, we have, in the unormalized case, 
\begin{equation}
\left( \langle k|+\langle k^{\prime }|\right) \hat{N}\left( |k\rangle
+|k^{\prime }\rangle \right) =k+k^{\prime },  \label{kk1}
\end{equation}%
and for the normalized superpositions
\begin{equation}
\left( \frac{\langle k|+\langle k^{\prime }|}{\sqrt{2}}\right) \hat{N}\left(
\frac{|k\rangle +|k^{\prime }\rangle }{\sqrt{2}}\right) =\frac{k+k^{\prime }%
}{2},  \label{kk2}
\end{equation}%
where $k+k^{\prime }$ is a sum of the integer numbers $k$ and $k^{\prime }$. 

Let us first consider non-normalized states. We proceed to the correspondence with the GCS 
defined by means of $C_{n-2}$-terms, for $n\geq 2$. If $n$ is not prime
\begin{eqnarray}
C_{n-2}&=& \lbrace 2n=p_{1}+p_{2} \rbrace, \label{1}
\end{eqnarray}
and if $n$ is a prime $n=p$
\begin{eqnarray}
C_{p-2}&=& \lbrace 2p=p_{1}+p_{2}= p+p \rbrace. \label{1}
\end{eqnarray}
The corresponding GCS in terms of expectation values of the number operator is described by means of the following sets of 
equations: If $n\geq 2$ is not a prime number, 
\begin{eqnarray}
\mathcal{Q}_{n-2}&=& \lbrace\langle 2n|\hat{N}|2n\rangle =\langle p_{1}|\hat{N}%
|p_{1}\rangle +\langle p_{2}|\hat{N}|p_{2}\rangle\rbrace.
\end{eqnarray}
and, if  $n\geq 2$ is a prime number $n=p$, 
\begin{eqnarray}
\mathcal{Q}_{p-2}&=& \lbrace\langle 2p|\hat{N}|2p\rangle =\langle p_{1}|\hat{N}%
|p_{1}\rangle +\langle p_{2}|\hat{N}|p_{2}\rangle 
\nonumber \\
&=& \langle p|\hat{N}%
|p\rangle +\langle p|\hat{N}|p\rangle\rbrace.
\end{eqnarray} 

We can check immediatelly that $C_{k}$ and $\mathcal{Q}_{k}$ correspond to equivalent GC terms 
for each $2(k+2)$ even number.

\section{Normalization of superpositions}
We can consider two possible types of superpositions of Fock states: a normalized
superposition (\ref{kk1}) or non-normalized superpostion (\ref{kk2}). Considering non-normalized states, we have derived 
$\mathcal{Q}_{n-2}$ equivalent to the GCS $C_{n-2}$. 

Now, let us consider the normalized case. The question of normalization rises an ambiguity, related to the equivalence of 
quantum states. We emphasize that far from trivial, the ambiguities in quantum states can be related to fundamental questions, as observed 
in order physical situations \cite{bala}.
 Let us start analysing specific cases.

Consider the state $|2\rangle$. The expectation value of $\hat{N}$ in this state is given simply by
$\langle 2|\hat{N}|2\rangle = 2$. On the other hand, the superposition $|2\rangle + |2\rangle = 2|2\rangle$ represents 
the same state as $|2\rangle$, with a denormalization factor $2$. But, if they correspond to the same physical state, the result 
of the expectation value should be the same for both states, i.e., $\langle 2|\hat{N}|2\rangle = 2$. Returning to 
the term $\mathcal{Q}_{0}$, the equality $\langle 4|\hat{N}|4\rangle = \langle 2|\hat{N}|2\rangle 
+\langle 2|\hat{N}|2\rangle=4$, we note that it is possible rewriting in terms of the following projection 
\begin{equation}
\langle 2|\hat{N}(|2\rangle +|2\rangle )= 4.
\end{equation}%
However, the state $|2\rangle +|2\rangle$ corresponds to the same state as $|2\rangle$, when we introduce the normalization, i.e.,
 $(|2\rangle +|2\rangle )/2$. By rewriting the superposition with this normalized superposition, we have
\begin{equation}
\langle 2|\hat{N}(\frac{|2\rangle +|2\rangle }{2})=\langle 2|\hat{N}%
|2\rangle=2.
\end{equation}
We can also take the expectation value
\begin{equation}
(\frac{\langle 2|+\langle 2|}{2})\hat{N}(\frac{|2\rangle +|2\rangle }{2})=2.
\end{equation}
Other important point is when the states are associated to two different prime numbers. Let us consider the specific case 
of $\mathcal{Q}_{2}$. When taking the normalization into account the sum $\langle 3|\hat{N}%
|3\rangle +\langle 5|\hat{N}|5\rangle$ can be considered as an expectation value in some superposition. Indeed, we have
\begin{eqnarray}
\left(\langle 3| \pm \langle 5|\right)\hat{N}\left(|3\rangle \pm |5\rangle \right)&=& \langle 3|\hat{N}%
|3\rangle +\langle 5|\hat{N}|5\rangle=8.\nonumber \\
\end{eqnarray}
In the normalized case, these superpositions have a $1/\sqrt{2}$ factor, such that
\begin{eqnarray}
\left(\frac{\langle 3| \pm \langle 5|}{\sqrt{2}}\right)\hat{N}\left(\frac{|3\rangle \pm |5\rangle}{\sqrt{2}} \right) &=& 4. 
\end{eqnarray}
We can generalize these results with a new sequence, considering normalized states. 

If $n$ is not a prime number, we can write 
\begin{eqnarray}
\widetilde{\mathcal{Q}}_{n-2} &=& \lbrace\langle 2n|\hat{N}|2n\rangle = 2n;\nonumber \\
&&\left(\frac{\langle p_{1}| \pm \langle p_{2}|}{\sqrt{2}}\right)\hat{N}\left(\frac{|p_{1}\rangle \pm |p_{2}\rangle}{\sqrt{2}} 
\right)= \frac{p_{1} + p_{2}}{2};\nonumber \\
&&\rbrace,
\end{eqnarray}
and if $n$ is a prime number $p$, we have
\begin{eqnarray}
\widetilde{\mathcal{Q}}_{p-2}&=& \lbrace\langle 2p|\hat{N}|2p\rangle = 2p;\nonumber \\
&&\left(\frac{\langle p_{1}| \pm \langle p_{2}|}{\sqrt{2}}\right)\hat{N}\left(\frac{|p_{2}
\rangle \pm |p_{1}\rangle}{\sqrt{2}} \right) = \frac{p_{1} + p_{2}}{2};  \nonumber \\
&& \left(\frac{\langle p|+\langle p|}{2}\right)\hat{N}\left(\frac{|p\rangle +|p\rangle }{2}\right)= p.\rbrace.
\end{eqnarray}

Thus, the normalization of superpositions will lead to a new sequence $\widetilde{\mathcal{Q}}_{n-2}$ that is 
not equivalent to the GCS $C_{n-2}$ and their equivalent $\mathcal{Q}_{n-2}$. 

\section{Algorithm for normalized GCS}

We can have an alternative to the normalized sequences that can be equivalent to $C_{n-2}$. We propose the following algorithm, associated to properties 
of Fock states: Consider an arbitrary Fock state $|m\rangle$. If $m>2$  is even, then we consider the expectation value in the 
number operator $\hat{N}$. If $m$ is odd and not prime, then we annihilate the state. Finally, if $m$ is a prime number, 
we consider the expectation value in the operator $2\hat{N}$, instead of $\hat{N}$, such that we arrive at the result
 $\langle m|2\hat{N}|m\rangle=2m$. For superpostions, we consider the two normalized superposition of two different Fock states 
$|m\rangle$ and $|m'\rangle$ given by $\left(|m\rangle + |m'\rangle\right)/\sqrt{2}$.
If any of these states is not prime we simply annihilate the state, such that we obtain only prime Fock state superpositions. We 
then consider the expectation values of the operator $2\hat{N}$, such that we will arrive at
\begin{eqnarray}
\left(\frac{\langle m| + \langle m'|}{\sqrt{2}}\right)2\hat{N}\left(\frac{|m\rangle + |m'\rangle}{\sqrt{2}}\right)= m + m'.
\end{eqnarray}
With this algorithm we can write a new sequence $\widetilde{\widetilde{\mathcal{Q}}}_{k}$ that will be equivalent to the 
GCS $C_{k}$, described by 
\begin{eqnarray}
\widetilde{\widetilde{\mathcal{Q}}}_{n-2}&=& \lbrace\langle 2n|\hat{N}|2n\rangle = 2n;\nonumber \\
&&\left(\frac{\langle p_{1}| \pm \langle p_{2}|}{\sqrt{2}}\right)2\hat{N}\left(\frac{|p_{1}\rangle \pm |p_{2}\rangle}{\sqrt{2}} 
\right)= p_{1} + p_{2} \rbrace, \nonumber \\
\end{eqnarray}
if $n$ is not a prime number
and by 
\begin{eqnarray}
\widetilde{\widetilde{\mathcal{Q}}}_{p-2}&=& \lbrace\langle 2p|\hat{N}|2p\rangle = 2p; \nonumber \\
&&\left(\frac{\langle p_{1}| \pm \langle p_{2}|}{\sqrt{2}}\right)\hat{N}\left(\frac{|p_{1}\rangle 
\pm |p_{2}\rangle}{\sqrt{2}} \right)= p_{1} + p_{2}; \nonumber \\
&& \left(\frac{\langle p|+\langle p|}{2}\right)2\hat{N}\left(\frac{|p\rangle +|p\rangle }{2}\right)= 2p.\rbrace,
\end{eqnarray}
if $n$ is a prime number $p$.

Notice that now we have an equivalence with the GCS $C_{n-2}$. In fact, 
the expectation values correspond to the tems in the GC, as $2n=2p$ or $2n=p_{1}+p_{2}$. The sequence 
$\widetilde{\widetilde{\mathcal{Q}}}_{n-2}$ is achieved by introducing the operator $2\hat{N}$ in 
the calculation of expectation values associated to superposition of prime numbers. It solves the problem 
with the factor $2$ that results of the equivalence between the states
\begin{equation}
\frac{|p_{1}\rangle \pm |p_{2}\rangle }{\sqrt{2}}\leftrightarrow \sqrt{2}%
\left( \frac{|p_{1}\rangle \pm |p_{2}\rangle }{\sqrt{2}}\right) ,
\end{equation}%
\begin{equation}
\frac{|p\rangle +|p\rangle }{2}\leftrightarrow 2\left( \frac{%
|p\rangle +|p\rangle }{2}\right).
\end{equation}
In this form, the operator for the sequence is in a symmetric form, such that when taking the expectation values
\begin{equation}
\left[ \left(
\frac{\langle p|+\langle p|}{2}\right) 2\hat{N}\left( \frac{%
|p\rangle +|p\rangle }{2}\right) \right],
\end{equation}
and
\begin{equation}
\left[ \left( \frac{\langle p_{1}|\pm \langle p_{2}|}{\sqrt{2}}\right)2\hat{N}\left( \frac{|p_{1}\rangle \pm |p_{2}\rangle }{\sqrt{2}}%
\right) \right],
\end{equation}
we achieve at the same results associated to the GCS. 

Since the Fock states are eigenstates of the following harmonic oscillator hamiltonian $\hat{H}=\omega\hbar\left(1/2 + \hat{N}\right)$,
the transformation from the number operator to the new operator, $\hat{N}\rightarrow 2\hat{N}$ corresponds 
to the following canonical transformation
\begin{eqnarray}
\hat{H}\rightarrow \hat{H}'= \omega\hbar\left(\frac{1}{2} + 2\hat{N}\right).
\end{eqnarray}
This new hamiltonian corresponds to an increasing in the energy eigenvalues and preserving the same vacuum fluctuations, i.e.,
$\hat{H}'|m\rangle=(\omega\hbar/2 +(2\omega)\hbar m)|m\rangle$. The part not associated to the vaccum fluctuation 
has an increasing in the frequency $\omega \rightarrow 2\omega$. The algorithm for the generation of 
$\widetilde{\widetilde{\mathcal{Q}}}_{n-2}$ then involves a canonical transformation procedure for the superposition of prime number states.

\section{Projections and states associated to GCS}

Let us return to the non-normalized sequence, the terms $\mathcal{Q}_{n-2}$. Taking into account 
the term $\mathcal{Q}_{0}$, we can write its corresponding equation in the following form
$\langle 4|\hat{N}|4\rangle -2\langle 2|\hat{N}|2\rangle =0$. Instead of trying to think in terms of expectation values, we can 
consider this equality as a projection involving possible superpositions of Fock states. Indeed, we can write that in the following 
way $\left( \langle 4|-2\langle 2|\right) \hat{N}(|4\rangle +|2\rangle )=0$.
Here the normalization is not a problem and we can include the normalization factors such that the equality can be written as
\begin{equation}
\left(\frac{\langle 4|-2\langle 2|}{\sqrt{5}}\right)\hat{N}\left(\frac{|4\rangle +|2\rangle }{\sqrt{2}}\right)=0.
\end{equation}
In this form, the GCS is associated to two states, whose projections followed by the action of the number operator 
correspond to an equivalent GC term. 

In order to consider the case for two different prime numbers, let us return to the $\mathcal{Q}_{2}$. We can write
$\langle 8|\hat{N}|8\rangle -\langle 5|\hat{N}|5\rangle -\langle 3|\hat{N}%
|3\rangle =0$. This equality can be written as a projection involving two different states
\begin{equation}
\left(\frac{\langle 8|-\langle 5|-\langle 3|}{\sqrt{3}}\right)\hat{N}%
\left(\frac{|8\rangle +|5\rangle +|3\rangle }{\sqrt{3}}\right)=0.
\end{equation}%
Since these projections do not lead to normalization problems, this is another alternative way to establish a 
 correspondence with the GCS $C_{k}$. We can then generalize the above for any $2n=p+p$ 
and $2n=p_{1}+p_{2}$, by defining the following states
\begin{eqnarray}
|\mathcal{G}_{2n,p}\rangle &=& \frac{|2n\rangle +|p\rangle }{\sqrt{2}}, \label{b123}\\
|\widetilde{\mathcal{G}}_{2n,p}\rangle&=&\frac{|2n\rangle -2|p\rangle }{\sqrt{5}},  \label{a123} \\
|\mathcal{G}_{2n,p_{1},p_{2}}^{(\pm)}\rangle&=& \frac{|2n\rangle \pm |p_{1}\rangle \pm |p_{2}\rangle }{%
\sqrt{3}}. \label{c123}
\end{eqnarray}
These states generate the GCS by means of the following projections
\begin{eqnarray}
&&\langle \mathcal{G}_{2n,p_{1},p_{2}}^{(\mp)}|\hat{N}|\mathcal{G}_{2n,p_{1},p_{2}}^{(\pm)}\rangle=0, \label{kdi12} \\
&&\langle \widetilde{\mathcal{G}}_{2n,p}|\hat{N}|\mathcal{G}_{2n,p}\rangle=0. \label{kdi13}
\end{eqnarray}
We can then write a new sequence equivalent to the GCS $C_{n-2}$, with the following 
sets of states and projections:  
\begin{eqnarray}
{\bf {\mathcal{G}}}_{n-2} &=& \lbrace |\mathcal{G}_{2n,p_{1},p_{2}}^{(\pm)}\rangle; 
 \langle \mathcal{G}_{2n,p_{1},p_{2}}^{(\mp)}|\hat{N}|\mathcal{G}_{2n,p_{1},p_{2}}^{(\pm)}\rangle=0. \rbrace, \nonumber \\
\end{eqnarray}
if $n$ is not a prime number and
\begin{eqnarray}
{\bf {\mathcal{G}}}_{p-2} &=& \lbrace |\mathcal{G}_{2p,p}\rangle; |\widetilde{\mathcal{G}}_{2p,p}\rangle; 
|\mathcal{G}_{2p,p_{1},p_{2}}^{(\pm)}\rangle; \nonumber \\
&& \langle \widetilde{\mathcal{G}}_{2p,p}|\hat{N}|\mathcal{G}_{2p,p}\rangle=0;  \nonumber \\
&& \langle \mathcal{G}_{2p,p_{1},p_{2}}^{(\mp)}|\hat{N}|\mathcal{G}_{2p,p_{1},p_{2}}^{(\pm)}\rangle=0. \rbrace,
\end{eqnarray}
if $n$ is a prime number $p$.

We can associate these states to states of harmonic oscillator by the 
following relations
\begin{eqnarray}
&&\frac{\sqrt{5}}{3}|\mathcal{G}_{2n,p}\rangle + \frac{2\sqrt{2}}{3}|\widetilde{\mathcal{G}}_{2n,p}\rangle = |2n\rangle, \label{1w} \\
&&\frac{\sqrt{2}}{3}|\mathcal{G}_{2n,p}\rangle - \frac{\sqrt{5}}{3}|\widetilde{\mathcal{G}}_{2n,p}\rangle = |p\rangle, \label{2w} \\
&& \frac{\sqrt{3}}{2}\left(|\mathcal{G}_{2n,p_{1},p_{2}}^{(+)}\rangle - |\mathcal{G}_{2n,p_{1},p_{2}}^{(-)}\rangle\right) \nonumber \\
&-&\left(\frac{\sqrt{2}}{3}|\mathcal{G}_{2p_{1},p_{1}}\rangle - \frac{\sqrt{5}}{3}|\widetilde{\mathcal{G}}_{2p_{1},p_{1}}\rangle\right)
=|p_{2}\rangle, 
\end{eqnarray}
and
\begin{eqnarray}
&& \frac{\sqrt{3}}{2}\left(|\mathcal{G}_{2n,p_{1},p_{2}}^{(+)}\rangle - |\mathcal{G}_{2n,p_{1},p_{2}}^{(-)}\rangle\right) \nonumber \\
&-&\left(\frac{\sqrt{2}}{3}|\mathcal{G}_{2p_{2},p_{2}}\rangle - \frac{\sqrt{5}}{3}|\widetilde{\mathcal{G}}_{2p_{2},p_{2}}\rangle\right)
=|p_{1}\rangle, 
\end{eqnarray}
It is immediate to show that these superpositions are normalized and correspond to eigenstates of the 
harmonic oscillator with eigenvalues 
$\omega \hbar\left(\frac{1}{2} + 2n\right)$, $\omega \hbar\left(\frac{1}{2} + p\right)$, $\omega \hbar\left(\frac{1}{2} + p_{1}\right)$
 and $\omega \hbar\left(\frac{1}{2} + p_{2}\right)$. 

We can then rewrite the action of the hamiltonian in the following form
\begin{eqnarray}
\hat{H}|\mathcal{G}_{2n,p}\rangle &=& \frac{\omega \hbar\left(\frac{1}{2} + 2n\right)}{\sqrt{2}}\left[\frac{\sqrt{5}}{3}|\mathcal{G}_{2n,p}\rangle + \frac{2\sqrt{2}}{3}|\widetilde{\mathcal{G}}_{2n,p}\rangle\right] \nonumber \\
&+& \frac{\omega \hbar\left(\frac{1}{2} + p\right)}{\sqrt{2}}\left[\frac{\sqrt{2}}{3}|\mathcal{G}_{2n,p}\rangle - \frac{\sqrt{5}}{3}|\widetilde{\mathcal{G}}_{2n,p}\rangle\right],  \nonumber \\ \label{b123} \\
\hat{H}|\widetilde{\mathcal{G}}_{2n,p}\rangle &=& \frac{\omega \hbar\left(\frac{1}{2} + 2n\right)}{\sqrt{5}}\left[\frac{\sqrt{5}}{3}|\mathcal{G}_{2n,p}\rangle + \frac{2\sqrt{2}}{3}|\widetilde{\mathcal{G}}_{2n,p}\rangle\right] \nonumber \\
&-& \frac{2\omega \hbar\left(\frac{1}{2} + p\right)}{\sqrt{5}}\left[\frac{\sqrt{2}}{3}|\mathcal{G}_{2n,p}\rangle - \frac{\sqrt{5}}{3}|\widetilde{\mathcal{G}}_{2n,p}\rangle\right],  \nonumber \\ \label{a123} \\
\hat{H}|\mathcal{G}_{2n,p_{1},p_{2}}^{(\pm)}\rangle&=& \frac{\omega \hbar\left(\frac{1}{2} + 2n\right)}{\sqrt{3}}\left[\frac{\sqrt{5}}{3}|\mathcal{G}_{2n,p}\rangle + \frac{2\sqrt{2}}{3}|\widetilde{\mathcal{G}}_{2n,p}\rangle\right] \nonumber \\
&\pm& \frac{\omega\hbar(p_{1}+p_{2} +1)}{2}\left[|\mathcal{G}_{2n,p_{1},p_{2}}^{(+)}\rangle 
- |\mathcal{G}_{2n,p_{1},p_{2}}^{(-)}\rangle\right] \nonumber \\
&\mp& \frac{\omega \hbar\left(\frac{1}{2} + p_{1}\right)}{\sqrt{3}}\left[\frac{\sqrt{2}}{3}|\mathcal{G}_{2p_{2},p_{2}}\rangle
 - \frac{\sqrt{5}}{3}|\widetilde{\mathcal{G}}_{2p_{2},p_{2}}\rangle\right] \nonumber 
\\
&\mp& \frac{\omega \hbar\left(\frac{1}{2} + p_{2}\right)}{\sqrt{3}}\left[\frac{\sqrt{2}}{3}|\mathcal{G}_{2p_{1},p_{1}}\rangle
 - \frac{\sqrt{5}}{3}|\widetilde{\mathcal{G}}_{2p_{1},p_{1}}\rangle\right]. \nonumber 
\\
\end{eqnarray}
The sequences ${\bf {\mathcal{G}}}_{n-2}$ are then associated to states of excitations in the harmonic oscillator and are 
equivalent to the GCS $C_{n-2}$.

\section{GC and spectrum of oscillator}

We can also consider the states $|2n\rangle, |2p\rangle, |p_{1}+p_{2}\rangle$ and  $|2n -p_{1} -p_{2}\rangle$, that
are associated as component terms in GCS. These states have particular eigenstates in an harmonic oscilator
\begin{eqnarray}
\hat{H}|2n\rangle&=&\omega\hbar\left(\frac{1}{2} + 2n\right)|2n\rangle, \\
\hat{H}|2p\rangle&=&\omega\hbar\left(\frac{1}{2} + 2p\right)|2p\rangle, \\
\hat{H}|p_{1}+p_{2}\rangle&=&\omega\hbar\left(\frac{1}{2} + p_{1}+p_{2}\right)|p_{1}+p_{2}\rangle, 
\end{eqnarray}
and
\begin{equation}
\hat{H}|2n -p_{1} -p_{2}\rangle=\omega\hbar[\frac{1}{2} + 2n -(p_{1}+p_{2}) ]|2n - (p_{1} + p_{2})\rangle. 
\end{equation}
As a consequence the GC terms are associated to the same spectrum in the hamiltonian. In the case of the state
$|2n -p_{1} -p_{2}\rangle$, the GC will be associated to the vacuum state $|0\rangle$, as such their energies are vacuum fluctuations.

\section{GC problem, Goldbach partition and Degeneracy of states}

\begin{figure}[h]
\centering
\includegraphics[scale=0.5]{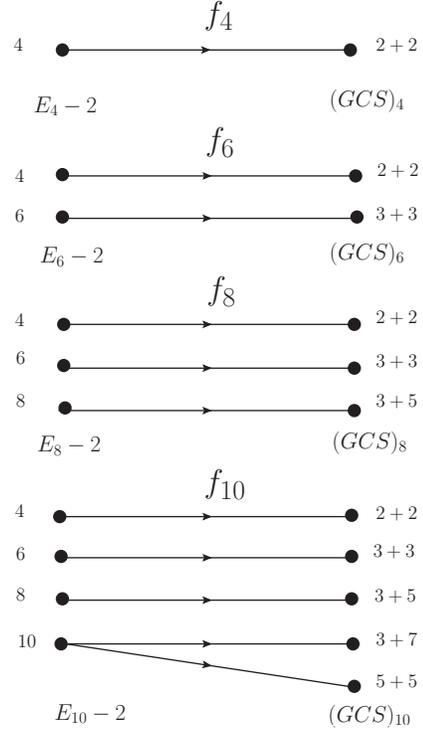}
\caption{Onto maps $f_{4}$, $f_{6}$, $f_{8}$ and $f_{10}$.} 
\label{c21}
\end{figure}

\begin{figure}[h]
\centering
\includegraphics[scale=0.5]{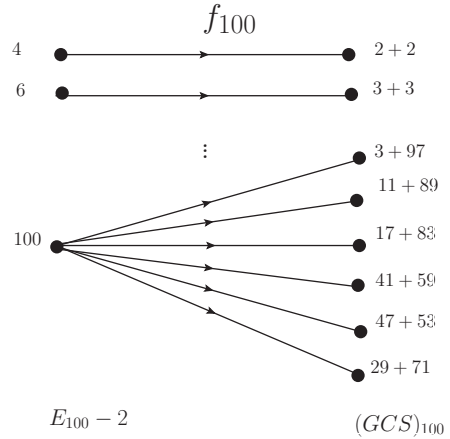}
\caption{Onto map $f_{100}$.} 
\label{c100}
\end{figure}

The GC problem can be seen as a question about mapping. We can divide positive integers in even numbers $2n$ and odd numbers 
$2n+1$. Considering an even number $2n$, if $n$ is prime, then we just multiply by $p$ and get the GCS $2n=2p= p+p$, as discussed 
above. If $n$ is not prime, $p_{1}$ and $p_{2}$ are two different prime numbers, 
the form $2n=p_{1} + p_{2}$ is applied. This discussion can also be started from the prime numbers. Let $p_{1}$ and $p_{2}$ 
be odd prime numbers, i. e., different from $2$. Their sum then is even, consequently
\begin{eqnarray}
p_{1} + p_{2}=2n.
\end{eqnarray}
On the other hand, let one of the numbers be $2$, then the sum is odd, 
\begin{eqnarray}
2 + p_{1} = 2n+1.
\end{eqnarray}
Consequently, the set of the numbers in the GC is a subset of even numbers. Let us call the set of even numbers
$E(n)=\lbrace 2n; n\in N \rbrace$. The set of GCS will be given by given by
\begin{eqnarray}
O_{\mathcal{P}}=\lbrace p_{1}+p_{2}; p_{1},p_{2}\in \mathcal{P}-2\rbrace \bigcup \lbrace 2 + 2 \rbrace.
\end{eqnarray}
Since all these sums are even, we have
\begin{eqnarray}
O_{\mathcal{P}}\subset E(n)-2,
\end{eqnarray}
where $E(n)-2$ is the set $E(n)$ without the number $2$.
But if the GC goes to all even numbers, we have to prove there is a surjective (onto) map
\begin{eqnarray}
f_{GC}: O_{\mathcal{P}} \rightarrow E(n)-2.
\end{eqnarray}
This accounts for the correspondence to generate all GCS. Although this is a problem in the infinity case, for a finite GCS it can 
be completely constructed for an arbitrarily large $n$. Let $E_{2n}-2$ be 
a set with even numbers until $2n$, excluding $2$, i. e., $n>1$. Let $(GCS)_{2n}$ be 
a finite sequence of GC terms until $2n=p_{1}+p_{2}$, $n>1$. As an example,
\begin{eqnarray}
E_{4}-2 &=&\lbrace 4\rbrace \\
E_{6}-2 &=&\lbrace 4,6\rbrace \\
E_{8}-2 &=&\lbrace 4,6,8\rbrace \\
&\vdots& \nonumber \\
E_{2n}-2 &=&\lbrace 4,6,8,..., 2n\rbrace.
\end{eqnarray}
The Goldbach partition imply that a given even number can have different ways to be written as a sum of prime numbers. 
We can define the following sets 
\begin{eqnarray}
(GCS)_{4} &=&\lbrace 4=2+2\rbrace \\
(GCS)_{6} &=&\lbrace 4=2+2,6=3+3\rbrace \\
(GCS)_{8} &=&\lbrace 4=2+2,6=3+3,8=3+5\rbrace \\
&\vdots& \nonumber \\
(GCS)_{2n} &=&\lbrace 4=2+2,6=3+3,..., 2n=p_{1}+p_{2}\rbrace. \nonumber \\
\end{eqnarray}
The onto maps $f_{2n}$ relating the sets $E_{2n}-2$ and $(GCS)_{2n}$, 
$f_{2n}: E_{2n}-2 \rightarrow (GCS)_{2n}$. Examples of these maps are showed in the figures \ref{c21} and \ref{c100}.
\begin{widetext}
\begin{figure}[h]
\centering
\includegraphics[scale=0.55]{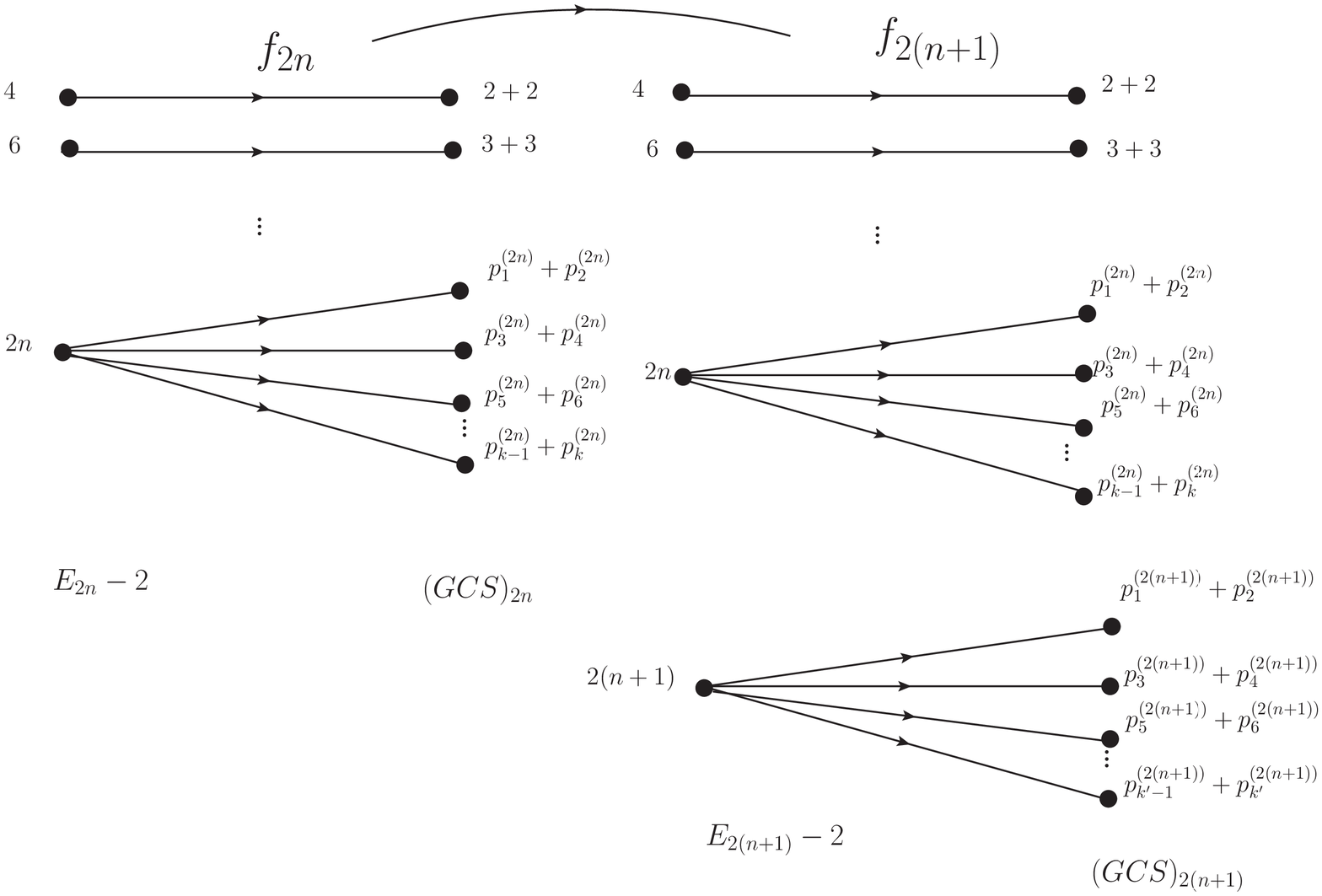}
\caption{Relation between the maps $f_{2n}$ and $f_{2(n+1)}$.} 
\label{c123}
\end{figure} 

The onto functions $f_{2n}$ and $f_{2(n+1)}$ are related by the fact that the GC term $2(n+1)$ is included in $f_{2n}$ to 
generate $f_{2(n+1)}$. As such, the onto functions can be built recursivelly from $2n$ to $2(n+1)$, as showed in the 
figure \ref{c123}.   

The sets $C_{k-2}$ associated to the GC tersm are related to the GCS sets $(GCS)_{2n}$ by means of the following relation 
\begin{eqnarray}
(GCS)_{2n}=\bigcup_{k=2}^{n}C_{k-2}. 
\end{eqnarray}
Due the equivalence with the sets associated to quantum states, we can derive equivalent onto maps:  
\begin{eqnarray}
f_{2n}^{\mathcal{Q}}: E_{2n}-2 \rightarrow \bigcup_{k=2}^{n}\mathcal{Q}_{k-2},
\end{eqnarray}
\begin{eqnarray}
f_{2n}^{\widetilde{\widetilde{\mathcal{Q}}}}: E_{2n}-2 \rightarrow \bigcup_{k=2}^{n}\widetilde{\widetilde{\mathcal{Q}}}_{k-2},
\end{eqnarray}
\begin{eqnarray}
f_{2n}^{{\bf {\mathcal{G}}}}: E_{2n}-2 \rightarrow \bigcup_{k=2}^{n}{\bf {\mathcal{G}}}_{k-2}.
\end{eqnarray}
\end{widetext}

\section{Conclusion}

We have demonstrated that a sequence 
$\mathcal{Q}_{k}$ in terms of expectation values of number operators in Fock states has a direct correspondence
to the GCS $C_{k}$. On the other hand, taking into account the normalization of Fock states, we have generated 
a new sequence $\widetilde{\mathcal{Q}}_{k}$ that is no more equivaltent to $C_{k}$. We propose an algorithm based 
on properties of Fock states, leading to a sequence 
$\widetilde{\widetilde{\mathcal{Q}}}_{k}$ in terms of expectation values, involving only normalized states and equivalent to the 
GCS $C_{k}$. We showed that this algorithm, involving 
a canonical transformation in the hamiltonian, solves the previous problem of factor $2$ in the sequence 
$\widetilde{\mathcal{Q}}_{k}$. We also proposed an alternative procedure, by deriving states associated to the GCS making use of projection 
relations. In this form, we showed that the correspondence to the GCS can also be realized by means of defined quantum states 
$|\mathcal{G}_{2n,p}\rangle$ and $|\mathcal{G}_{2n,p_{1},p_{2}}^{(\pm)}\rangle$, without problems with normalization. 
We derived relations of these states with eigenstates of an harmonic oscillator and the action of the corresponding hamiltonian. We also have showed that Fock states 
$|2n\rangle$, $|2p\rangle$, $|p_{1}+p_{2}\rangle$ and $|2n-p_{1}-p_{2}\rangle$ directly relate 
the GCS to the spectrum an harmonic oscillator. Finally, we addressed the problem of degeneracy associated to Goldbach partitions and discussed onto maps that build GCS at arbitrary 
size. Using the equivalence with the sets relating quantum states, we also showed that such maps can be established using quantum 
states. This possibility implies that GCS can be generated in different physical implementations, in particular cases where Fock states 
can be completelly controlled, as quantum optical systems.

\section{Acknowledgements}

The authors acknowledge B. Baseia, D. Bazeia and M. M. Ferreira Jr. for reading 
and suggestions on previous versions of the manuscript. %and the 
%referees for valuable suggestions that made this manuscript more improved. 
E. O. Silva acknowledges CNPq and FAPEMA 
for finantial support.


\begin{thebibliography}{99}
\bibitem{yuan} Y. Wang, \textit{The Goldbach Conjecture} (World Scientific,
2002). 
\bibitem{dox} A. Doxiadis, \textit{Uncle Petros and Goldbach's Conjecture} (Bloomsbury, 2001).
\bibitem{languasco} A. Languasco, Rend. Sem. Mat. Unv. Pol. Torino \textbf{53} (1995) 4.
\bibitem{sierra} G. Sierra, J. Rodriguez-Laguna, Phys. Rev. Lett. {\bf 106} (2011) 200201. 
\bibitem{cantor} G. Cantor, \textit{Contributions to the Founding of the Theory of Transfinite Numbers} (Dover, 1915).
\bibitem{shor} P. W. Shor, SIAM J. Comput. 26, 1484 (1997).
\bibitem{sierra2} J. I. Latorre, G. Sierra, arXiv:1302.6245.
\bibitem{aq} K. S. Gupta, E. Harikumar, A. R. de Queiroz, Eur. Phys. Lett. {\bf 102} (2013) 10006. 
\bibitem{lidar} A. T. Rezakhani et al., Phys. Rev. Lett. {\bf 103} (2009) 080502.
\bibitem{lozano} M. A. Sanchis-Lozano, J. F. Barbero, J. Navarro-Salas, Int. J. Modern Physics A, {\bf 27} (2012) 
1250136.
\bibitem{lu} N. Lu, Phys. Rev. A, \textbf{40} (1989) 1707.
\bibitem{bala} A. P. Balachandran, T. R. Govindarajan, A. R. de Queiroz, A. F. Reyes-Lega, Phys. Rev. Lett. {\bf 110} (2013) 080503. 
\bibitem{hofheinz} M. Hofheinz, et al. Nature, \textbf{454} (2008) 310.
\bibitem{wang} H. Wang et al., Phys. Rev. Lett. \textbf{101} (2008) 240401.
\bibitem{brune1} M. Brune et al., Phys. Rev. Lett., \textbf{101} (2008)
240402.
\bibitem{brune2} M. Brune, S. Haroche, V. Lefevre, J.M. Raimond and N.
Zagury, Phys. Rev. Lett., 65, 976 (1990).
\bibitem{raimond} J. M. Raimond, M. Brune, S. Haroche, Rev. Mod. Phys.,
\textbf{73} (2001) 565.
\bibitem{ralph} T.C.Ralph et. al., Phys. Rev. A, \textbf{73} (2006) 012113.
\bibitem{lupascu} A. Lupascu et al., Nature Physics, \textbf{3} (2007) 119.
\bibitem{sayrin} C. Sayrin, et al., Nature, \textbf{477} (2011) 73.
\bibitem{guerlin} C. Guerlin et al., Nature, \textbf{448} (2007) 889.
\bibitem{chun} C. S. Chuu et al., Phys. Rev. Lett. \textbf{95} (2005) 260403.
\bibitem{pons} M. Pons et al., Phys. Rev. A, \textbf{79} (2009) 033629.
\bibitem{ss} D. Sokolovski, M. Pons, A. del Campo, J. G. Muga, Phys. Rev. A,
\textbf{83} (2011) 013402.
\end{thebibliography}
\end{document}